\documentclass[twocolumn,prb,superscriptaddress,floatfix,showpacs,reprint,10pt]{revtex4-1}
\usepackage{graphicx,amsmath,bm}
\usepackage{times}

\def\<{\langle}
\def\>{\rangle}
\renewcommand{\vec}[1]{{\bf #1}}
\def\PNAS{Proc.~Natl.~Acad.~Sci.~U.S.A.~}
\def\PRL{Phys.~Rev.~Lett.~}
\def\PRB{Phys.~Rev.~B~}
\def\RMP{Rev.~Mod.~Phys.~}
\begin{document}
\title{Disorder and quasiparticle interference in heavy-fermion materials}
\author{Francesco~\surname{Parisen~Toldin}}
\affiliation{Max-Planck-Institute f\"ur Physik komplexer Systeme, N\"othnitzer Str.~38, D-01187 Dresden, Germany}
\author{Jeremy Figgins}
\affiliation{Department of Physics, University of Illinois at Chicago, Chicago, Illinois 60607, USA}
\author{Stefan Kirchner}
\affiliation{Max-Planck-Institut f\"ur Physik komplexer Systeme, N\"othnitzer Str.~38, D-01187 Dresden, Germany}
\affiliation{Max-Planck-Institut f\"ur Chemische Physik fester Stoffe, N\"othnitzer Str.~40, D-01187 Dresden, Germany}
\author{Dirk~K.~\surname{Morr}}
\affiliation{Max-Planck-Institut f\"ur Physik komplexer Systeme, N\"othnitzer Str.~38, D-01187 Dresden, Germany}
\affiliation{Department of Physics, University of Illinois at Chicago, Chicago, Illinois 60607, USA}

\begin{abstract}
Using a large-$N$ approach, we study the effect of disorder in the Kondo-screened phase of heavy-fermion materials. We demonstrate that the strong feedback between the hybridization and the conduction electron charge density magnifies the effect of disorder, such that already small concentrations of defects strongly disorder the materials' local electronic structure, while only weakly affecting their spatially averaged, thermodynamic properties. Finally, we show that the microscopic nature of defects can be identified through their characteristic signatures in the hybridization and quasiparticle interference spectrum.
\end{abstract}

\maketitle

{\it Introduction.}
Heavy-fermion materials are characterized by the presence of localized degrees of freedom, i.e.,
magnetic moments residing on rare-earth or actinide ions and itinerant $spd-$electronic states and the strong correlations between these degrees of freedom.~\cite{Hewson} The resulting interplay gives rise to a wide range of ground states ranging
from magnetic and superconducting phases~\cite{Stockert,Thompson,Aoki}  to semiconducting and metallic phases~\cite{basicrefs} with strongly enhanced quasiparticle mass, or without well-defined quasiparticles,\cite{Pfau} and even to enigmatic phases with yet unknown order parameters.~\cite{Palstra} Only recently, scanning tunneling spectroscopy (STS) experiments have succeeded in probing the local electronic structure of
several heavy-fermion compounds, such as URu$_2$Si$_2$, \cite{Schmidt-10,Aynajian-10,Hamidian-11}
YbRh$_2$Si$_2$, \cite{Ernst-10} CeRhIn$_5$ and CeCoIn$_5$,~\cite{Aynajian-12,All13} at sufficiently low energy and temperature to infer ground-state properties. In particular, by utilizing the spatial oscillations in the differential conductance, $dI/dV$, induced by defects and performing a quasiparticle-interference (QPI) analysis, it was possible to map out the electronic band structure near the Fermi energy.~\cite{Schmidt-10, Aynajian-12, All13, Yuan-12} STS experiments also reported defect-induced spatial oscillations in the hybridization possessing a wavelength that is determined by the unhybridized, small Fermi surface of the conduction band, \cite{Hamidian-11} thus confirming an earlier prediction by two of us. \cite{FM-11} Surprisingly, the same STS experiments also found that already a small, 1\% concentration of defects strongly disorders the hybridization in the entire system, while thermodynamic bulk measurements are largely insensitive to doping levels up to a few percent.~\cite{Pikul} Resolving this apparent contradiction between STS and thermodynamics clearly requires a more microscopic understanding of defect-induced effects in heavy-fermion materials.

In this Rapid Communication, we address this issue by computing the effects of finite impurity concentrations on the electronic and magnetic properties of heavy-fermion materials. We show that a strong feedback between the defect-induced spatial oscillations in the hybridization and the charge density of the conduction band leads to significant disorder in the local electronic properties already for small impurity concentrations. At the same time, thermodynamic properties of the system, such as the specific heat, are only weakly affected by defect concentrations of a few percent, thus explaining the qualitatively different STS and thermodynamic observations. Finally, our self-consistent treatment reveals that the form of the hybridization oscillations and of the QPI spectrum varies for different types of impurities. This result is not only of great importance for the interpretation of ongoing STM experiments, \cite{Aynajian-12,All13} but can also be employed to gain insight into the microscopic nature of disorder.

{\it Model.}
To study the effects of defects in heavy-fermion materials, we consider the Kondo-Heisenberg Hamiltonian
\begin{equation}
\label{hamiltonian}
\begin{split}
{\cal H} = -t\sum_{\<{\bf r,r'}\>,\sigma} c^\dagger_{{\bf r},\sigma} c_{{\bf r'},\sigma} -\mu\sum_{{\bf r},\sigma} c^\dagger_{{\bf r},\sigma} c_{{\bf r},\sigma}\\
+ J {\sum_{{\bf r}}} {\bf S}^{K}_{\bf r} \cdot {\bf s}^c_{\bf r} + I{\sum_{\<{\bf r,r'}\>}} {\bf S}^{K}_{\bf r} \cdot {\bf S}^{K}_{\bf r'},
\end{split}
\end{equation}
where $t$ is the hopping element between nearest-neighbor sites in a two-dimensional square lattice describing the light conduction band, and $c^\dagger_{{\bf r},\sigma}$ ($c_{{\bf r},\sigma}$) creates (destroys) a conduction electron of spin $\sigma$ at site ${\bf r}$. We choose a chemical potential of $\mu=-3.618t$, resulting in a Fermi wavelength  $\lambda_F^c=10$ (we set the lattice spacing, $a_0$, to unity) and an electron density of $n_c\simeq0.062$ of the (decoupled) conduction band. $J>0$ is the Kondo coupling between the conduction electron spin operator, ${\bf s}^c_{\bf r}$, and ${\bf S}^{K}_{\bf r}$, the $S=1/2$ spin of the magnetic atoms, and $I$ is the strength of the nearest-neighbors antiferromagnetic interaction between the magnetic atoms, which is treated here as an independent coupling constant. \cite{Yuan-12} Moreover, we consider defects in the form of missing magnetic atoms (i.e., vacancies), and nonmagnetic atoms that are substituted for magnetic ones, where the latter lead to a potential scattering term
$U\sum_{\vec{R},\sigma}c^\dagger_{{\bf R},\sigma} c_{{\bf R},\sigma}$, at the sites $\vec{R}$ of the nonmagnetic defects.

In the large$-N$ approach, \cite{Read83,Read,PPN-07} ${\bf S}^{K}_{\bf r}$ is represented by pseudofermion operators $f_{{\bf r},\sigma}^\dag$, $f_{{\bf r},\sigma}$ whose occupation number $\hat{n}_f(\vec{r}) \equiv \sum_\sigma \langle f_{{\bf r},\sigma}^\dag f_{{\bf r},\sigma} \rangle $ obeys the constraint $\hat{n}_f(\vec{r})=N/2$, with $N=2$ for $S=1/2$.
By adding the term $\sum_{{\bf r},\alpha}
\varepsilon_f({\bf r}) f^\dagger_{{\bf r},\alpha} f_{{\bf
r},\alpha}$ to the Hamiltonian, the constraint $\langle {\hat
n}_f({\bf r}) \rangle = N/2$ can be enforced through the on-site
energy $\varepsilon_f({\bf r})$. \cite{Kaul07,Igl97}
The quartic interactions in Eq.~(\ref{hamiltonian}) can then be decoupled by introducing the mean
fields \cite{FM-11}
\begin{equation}
s({\bf r}) = \frac{J}{2}\sum_{\alpha} \langle f^\dagger_{{\bf
r},\alpha} c_{{\bf r},\alpha} \rangle \, ; \quad \chi({{\bf r,r'}})
= \frac{I}{2}\sum_{\alpha} \langle f^\dagger_{{\bf
r},\alpha} f_{{\bf r'},\alpha} \rangle \label{eq:2} \ .
\end{equation}
describing the local hybridization and the nearest-neighbors antiferromagnetic correlations, respectively. The resulting
quadratic Hamiltonian can be diagonalized in real space [assuming
periodic boundary conditions for an $(L \times L)$ lattice, where below, we take $L=41$],
allowing a self-consistent calculation of $s({\bf r})$, $\chi({{\bf
r,r'}})$, and $\varepsilon_f({\bf r})$. Below, we take for concreteness $J=2t$, $I=0.002t$, and temperature $T=0.00022t$ (unless otherwise stated) and thus study systems
well inside the Kondo-screened regime where $s({\bf r}) \not = 0$
for all sites and magnetic fluctuations and the resulting corrections to the mean-field
level are expected to be weak. \cite{PPN-07,Read83}

\begin{figure}[b]
\includegraphics[width=8.5cm]{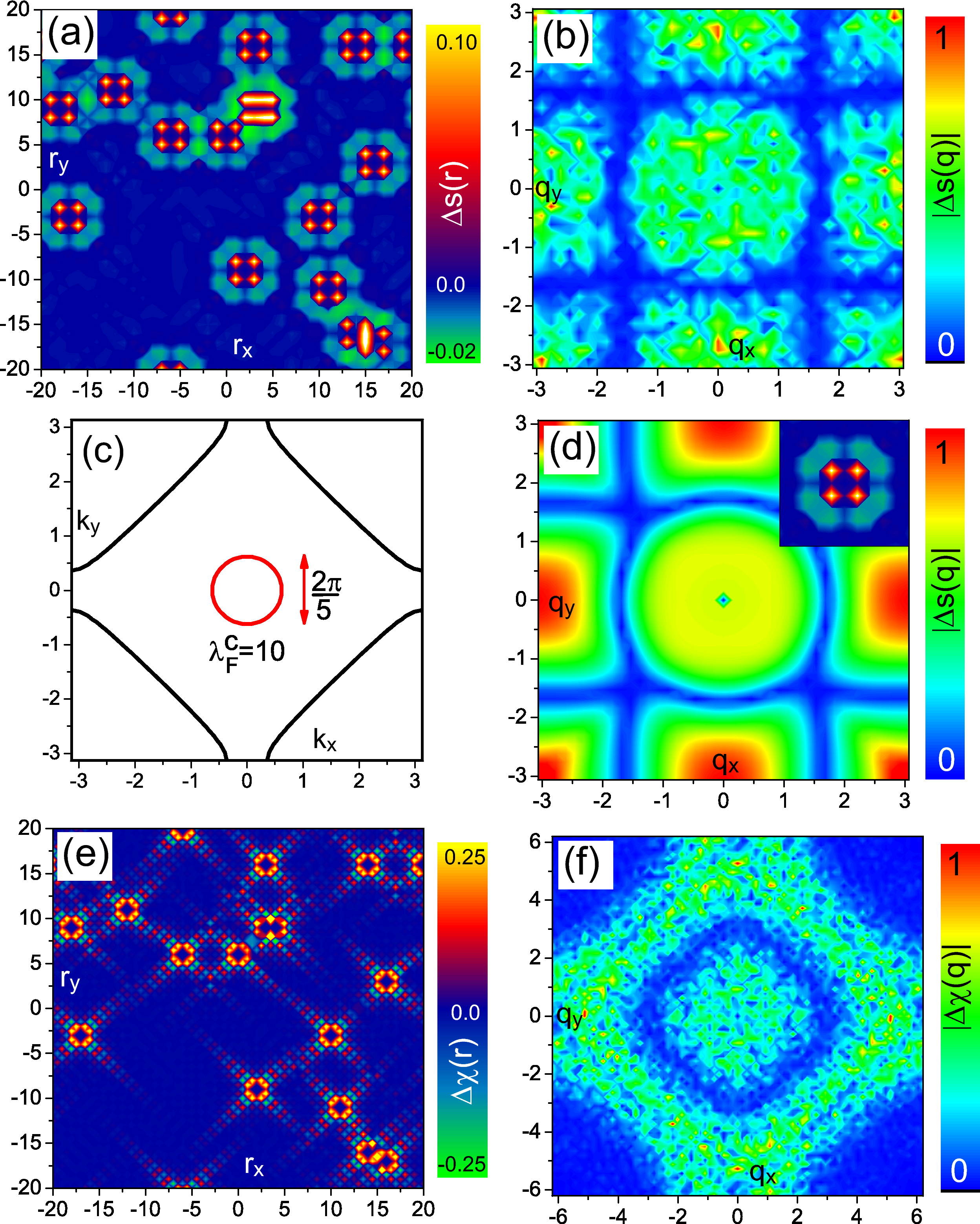}
\caption{(Color online) Kondo lattice with $1\%$ of vacancies. Contour plot of (a) $\Delta s(\bf{r})$, and (b) the absolute value of its Fourier transform, $|\Delta s(\bf{q})|$. (c) Large and small Fermi surfaces for an unperturbed lattice. (d) $|\Delta s(\bf{q})|$ and $\Delta s(\bf{r})$ (see inset) for a single vacancy. Contour plot of (e) $\Delta\chi({\bf r}_m)$ and (f) $|\Delta \chi(\bf{q})|$.}
\label{fig:j2u0}
\end{figure}

{\it Results.}
We begin by considering a system with 17 vacancies \cite{Kaul07} ($\sim 1\%$ concentration of defects) and present in Figs.~\ref{fig:j2u0}(a) and \ref{fig:j2u0}(b) the relative variation of the hybridization $\Delta s(\vec{r}) = (s(\vec{r}) - s_0)/s_0$, where $s_0$ is the hybridization of a clean lattice, and the corresponding absolute value of its Fourier transform (FT), $|\Delta s({\bf q})|$, respectively. As predicted earlier \cite{FM-11} and recently confirmed by STS experiments on URu$_2$Si$_2$ \cite{Hamidian-11}, $\Delta s(\vec{r})$ exhibits isotropic oscillations with wavelength $\lambda_s=\lambda_F^c/2=5$ [see Fig.~\ref{fig:j2u0}(a)], arising from the Fermi surface of the unhybridized  conduction band [see Fig.~\ref{fig:j2u0}(c)].

To understand the general momentum dependence of $|\Delta s({\bf q})|$, and the effects of random disorder, we present in Fig.~\ref{fig:j2u0}(d) $|\Delta s({\bf q})|$ for a single vacancy. A comparison of $|\Delta s({\bf q})|$ in Figs.~\ref{fig:j2u0}(b) and \ref{fig:j2u0}(d) shows that a finite concentration of defects leads as expected to a less well-defined momentum structure of $|\Delta s({\bf q})|$. Moreover, a detailed analysis shows that the momentum dependence of $|\Delta s({\bf q})|$, and in particular its maxima at $q=(\pi,\pi)$, $q=(0,\pi)$, and $q=(\pi,0)$, are predominantly determined by the four strong peaks in $\Delta s({\bf r})$ in the immediate vicinity of the vacancy [see inset of Fig.~\ref{fig:j2u0}(d)]. In $|\Delta s({\bf q})|$, these peaks completely overshadow the long-distance, $\lambda_F^c/2$ oscillations in $\Delta s(\vec{r})$ (whose amplitude is much smaller) such that their expected signature in $|\Delta s({\bf q})|$ at $q_s=2 \pi /\lambda_s \approx 1.26 $ is not clearly observed [see Fig.~\ref{fig:j2u0}(b)].
Finally, we present in Fig.~\ref{fig:j2u0}(e) the relative variation of the magnetic bond variable
$\Delta \chi(\bf{r}, \bf{r'})\equiv (\chi(\bf{r}, \bf{r'}) - \chi_{0})/\chi_0$ [plotted at ${\bf r}_m = (\bf{r} + \bf{r'})/2$]  where $\chi_0$ is the magnetic bond variable for a clean  lattice. As before, \cite{FM-11} we find that $\Delta \chi(\bf{r}, \bf{r'})$ exhibits strong oscillations along the lattice diagonal, arising from the large degree of nesting of the hybridized FS. \cite{FM-11} As a result, the absolute value of the Fourier transform $|\Delta \chi(\bf{q}_m)|$ of $\Delta \chi(\bf{r}_m)$ shown in Fig.~\ref{fig:j2u0}(f) exhibits the anisotropic form of the hybridized FS [see Fig.~\ref{fig:j2u0}(c)] [note that $\Delta \chi(\bf{r}_m)$ possesses a lattice constant of $a_0/2$, implying that for its first Brillouin zone, one has  $-2\pi/a_0 \leq q_m^{(x,y)} \leq 2\pi/a_0$].

\begin{figure}[t]
\includegraphics[width=8.5cm]{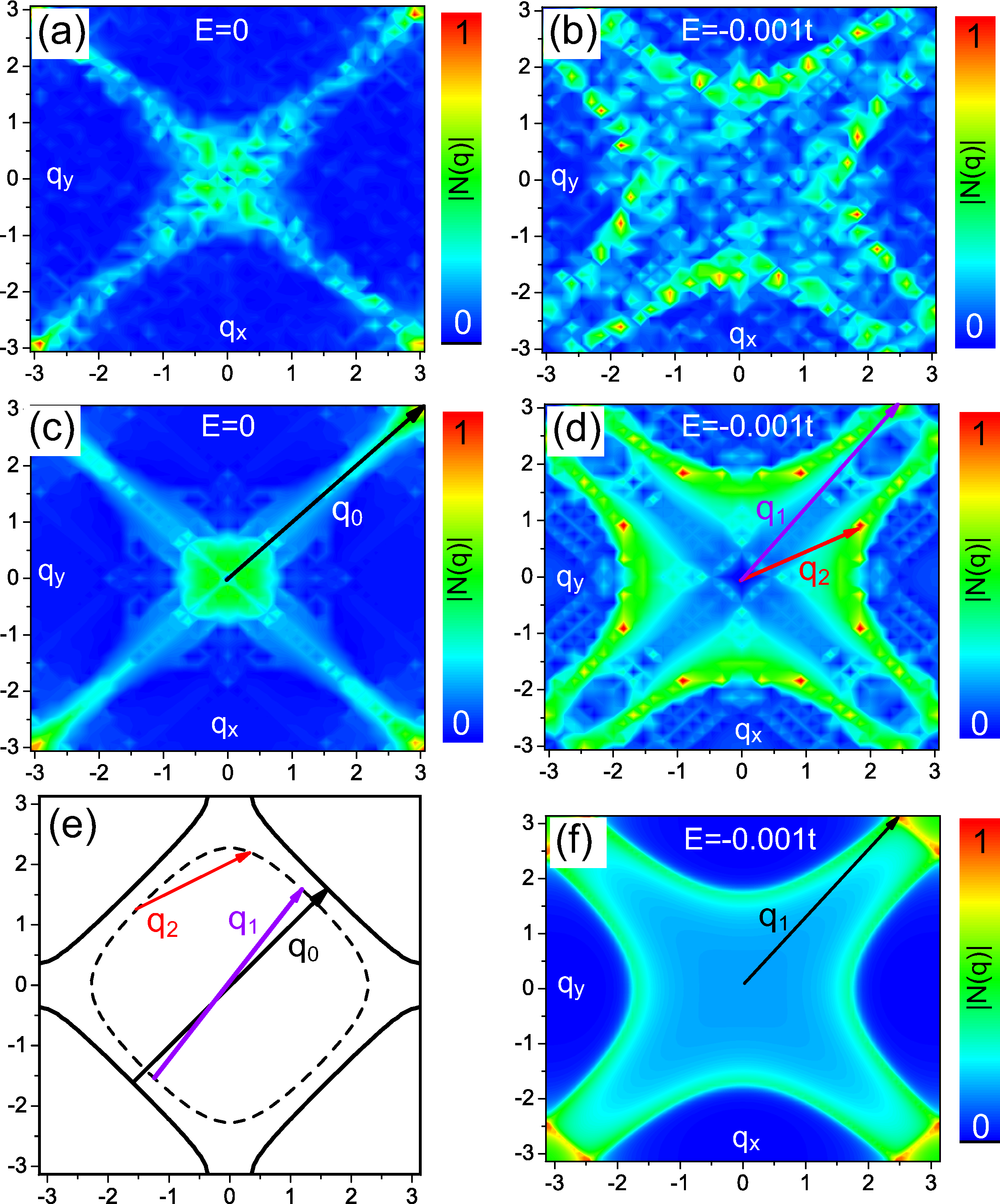}
\caption{(Color online) Contour plot of the QPI intensity $|N({\bf q})|$ for (a), (b) the system of Fig.~\ref{fig:j2u0}, and (c), (d) a system with a single defect, for $E=0$ and $E=-0.001t$. (e) Scattering vectors dominating the QPI intensity and equal energy contours. (f) QPI intensity obtained using the Born approximation. }
\label{fig:QPIJ2U0}
\end{figure}

We next consider the differential conductance $dI/dV$ and the absolute value of its FT, the QPI intensity $|N({\bf q})|$
(for a derivation, see Ref.~\cite{Yuan-12}). It was previously shown  \cite{FM-10} that the energy-dependent $dI/dV$ lineshape sensitively depends on the ratio $t_f/t_c$ (which is of the order of a few percent), where $t_c$ and $t_f$ are the amplitudes for tunneling of electrons from the STS tip into the conduction and $f$-electron bands, respectively. While the overall magnitude of $|N({\bf q})|$ varies with $t_f/t_c$, we find that its momentum dependence is rather insensitive to $t_f/t_c$, and we therefore take for concreteness $t_f/t_c=0.03$. In Figs.~\ref{fig:QPIJ2U0}(a) and \ref{fig:QPIJ2U0}(b) we present the QPI spectrum, obtained for the system shown in Fig.~\ref{fig:j2u0}, at two different energies. The presence of randomly distributed defects smears out the QPI spectra, as follows from a comparison with those spectra obtained for a single defect, shown in Figs.~\ref{fig:QPIJ2U0}(c) and \ref{fig:QPIJ2U0}(d). At the Fermi level, $E=0$, the QPI spectrum exhibits a peak at ${\bf q}_0=(\pi,\pi)$ [see Fig.~\ref{fig:QPIJ2U0}(c)], arising from scattering between the almost parallel portions of the FS, as shown in Fig.~\ref{fig:QPIJ2U0}(e). With decreasing energy, new peaks appear in the QPI spectrum away from the diagonal [${\bf q}_2$ in Figs.~\ref{fig:QPIJ2U0}(b) and (d)] whose spectral weight quickly becomes larger than that of the peaks, denoted by ${\bf q}_1$, close to the diagonal [Fig.~\ref{fig:QPIJ2U0}(e)]. This shift is a direct consequence of the spatial variation of the hybridization, as follows from a comparison with a QPI spectrum obtained within the Born approximation \cite{Yuan-12} where the hybridization is spatially constant, and the main spectral weight still resides with the peaks  (at ${\bf q}_1$) close to the diagonal [see Fig.~\ref{fig:QPIJ2U0}(f)]. We thus conclude that the spatial oscillations in the hybridization affect the QPI spectra, and in particular their spectral weight distribution, which is of great importance for the interpretation of experimentally obtained spectra, and the extraction of the underlying electronic band structure.

\begin{figure}[t]
\includegraphics[width=8.5cm]{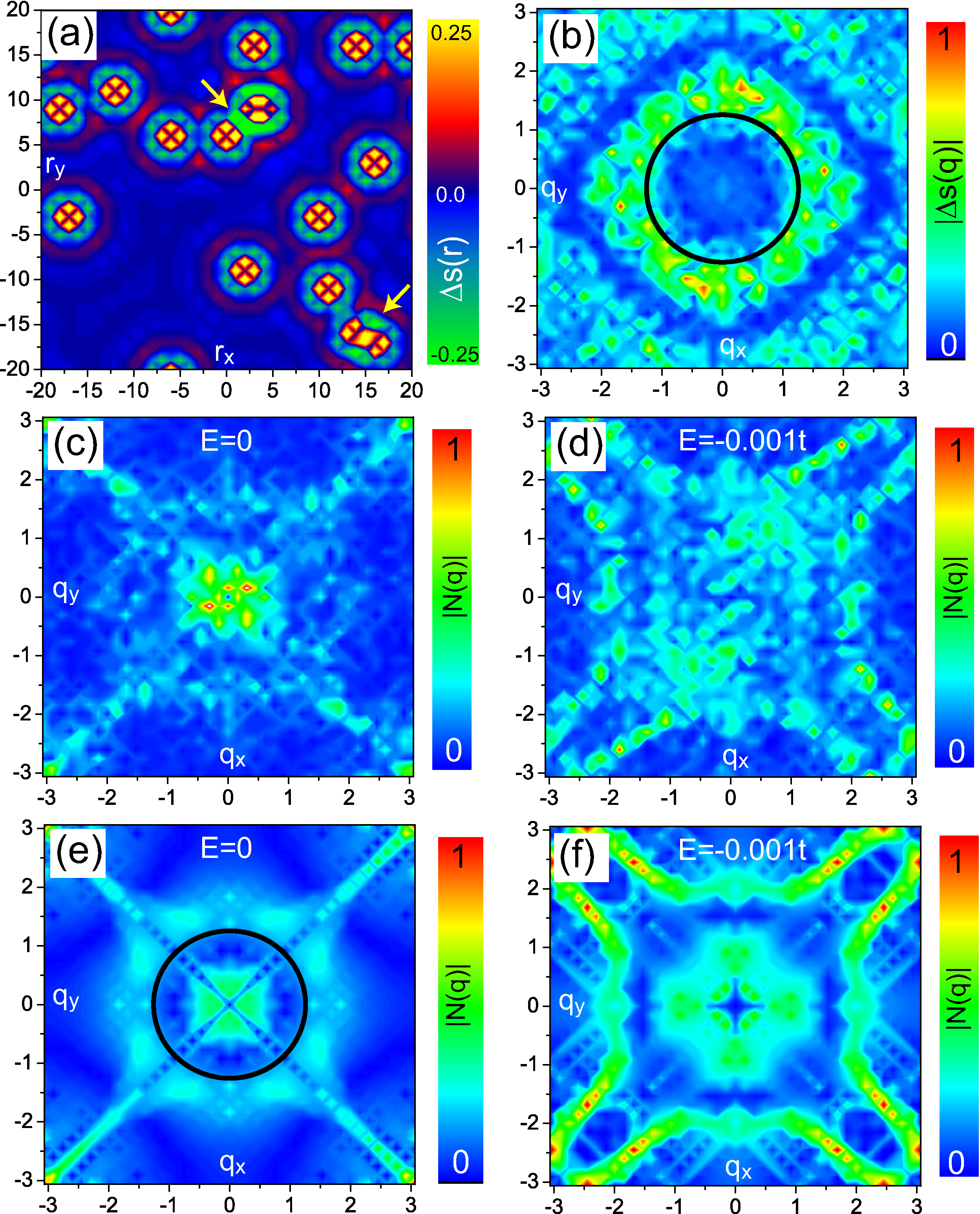}
\caption{(Color online) Kondo lattice with $1\%$ of nonmagnetic impurities with $U=-0.7t$. Contour plot of (a) $\Delta s(\bf{r})$, and (b) the absolute value of its Fourier transform, $|\Delta s(\bf{q})|$.  QPI intensity for a system with 1\% defects (c), (d) and with a single defect (e), (f).}
\label{fig:j2u-0.7}
\end{figure}

The above results change qualitatively when one considers a system with 1\% of nonmagnetic defects  (see Fig.~\ref{fig:j2u-0.7}) with scattering strength $U=-0.7t$, where the defects are located at the same positions as the vacancies of Fig.~\ref{fig:j2u0}.
A comparison of $\Delta s(\bf{r})$ shown in Fig.~\ref{fig:j2u-0.7}(a) with that in Fig.~\ref{fig:j2u0}(a) demonstrates that the introduction of a nonmagnetic scattering potential significantly alters the spatial pattern of hybridization oscillations. In particular, the oscillations acquire a larger amplitude, become much more isotropic, and the maxima in $\Delta s(\bf{r})$ in the immediate vicinity of the defect are rotated by $\pi/4$. These changes are particularly apparent in the Fourier transform, $|\Delta s(\bf{q})|$, shown in Fig.~\ref{fig:j2u-0.7}(b), which exhibits an almost isotropic pattern. These changes arise from a strong feedback effect of the charge density on the hybridization: The $s$-wave form of the nonmagnetic scattering potential leads to almost isotropic spatial oscillations of the conduction electron charge density \cite{FM-11} (not shown), which are reflected in those of $\Delta s(\bf{r})$.  As a result, the oscillations in $|\Delta s(\bf{q})|$ are dominated by $2k_F$ oscillations of the unhybridized Fermi surface [see black line in Fig.~\ref{fig:j2u-0.7}(b)] reflecting the strongly coupled nature of the system. The fact that the maximum intensity in $|\Delta s(\bf{q})|$ is located at slightly larger momenta than $2k_F$ arises from the exponential envelope of the spatial oscillations in $\Delta s$ \cite{FM-11} and the short decay length  ($\xi \approx 2.2$). Thus, while both vacancies and nonmagnetic defects lead to spatial oscillations in $\Delta s$ with wavelength $\lambda_s=\lambda_F^c/2$, these oscillations only become visible in the Fourier transform $\Delta s(\bf{q})$ if their amplitude is enhanced by oscillations in the conduction electron charge density induced by the nonmagnetic defects.
These conclusions provide insight into the microscopic nature of defects: In particular, the recent observation of hybridization oscillations in 1\% Th-doped URu$_2$Si$_2$, \cite{Hamidian-11} where the same $2k_F$ oscillations can be found in $\Delta s(\bf{r})$ and $\Delta s(\bf{q})$, suggests that the Th atoms exert an appreciable nonmagnetic scattering potential. The changes in $\Delta s(\bf{q})$ are also reflected in the QPI spectra shown in Figs.~\ref{fig:j2u-0.7}(c) and \ref{fig:j2u-0.7}(d) [for comparison, we also present in  Figs.~\ref{fig:j2u-0.7}(e) and \ref{fig:j2u-0.7}(f) the QPI spectra for a system with a single nonmagnetic defect], which now differ significantly from those obtained in a system with vacancies (see Fig.~\ref{fig:QPIJ2U0}). In particular, new peaks emerge in the QPI spectrum for $E=0$ [see black line in Fig.~\ref{fig:j2u-0.7}(e)] which reflect those peaks found in $|N({\bf q})|$ [see Fig.~\ref{fig:j2u-0.7}(b)]. We note that in the area between two (closely) placed defects [see yellow arrows in Fig.~\ref{fig:j2u-0.7}(a)] the spatial hybridization pattern is strongly affected by nonlinear quantum interference effects. In particular, comparing the local hybridization pattern with that obtained by simply superposing the hybridization patterns of single, noninterfering defects, we find that these nonlinearities in general suppress large oscillations in the hybridization, i.e., in $\Delta s({\bf r})$. Finally, we verified that the above results remain qualitatively unchanged for defect concentrations up to 2\% and different spatial disorder realizations.

Next, we demonstrate that even well below the Kondo temperature of the system, the spatial hybridization pattern and the QPI spectra can exhibit a significant temperature dependence. To this end, we again consider the system of Fig.~\ref{fig:j2u0} with 1\% of vacancies but with a smaller $J=1.47t$ and $I=0.0002t$.
\begin{figure}[t]
\includegraphics[width=8.5cm]{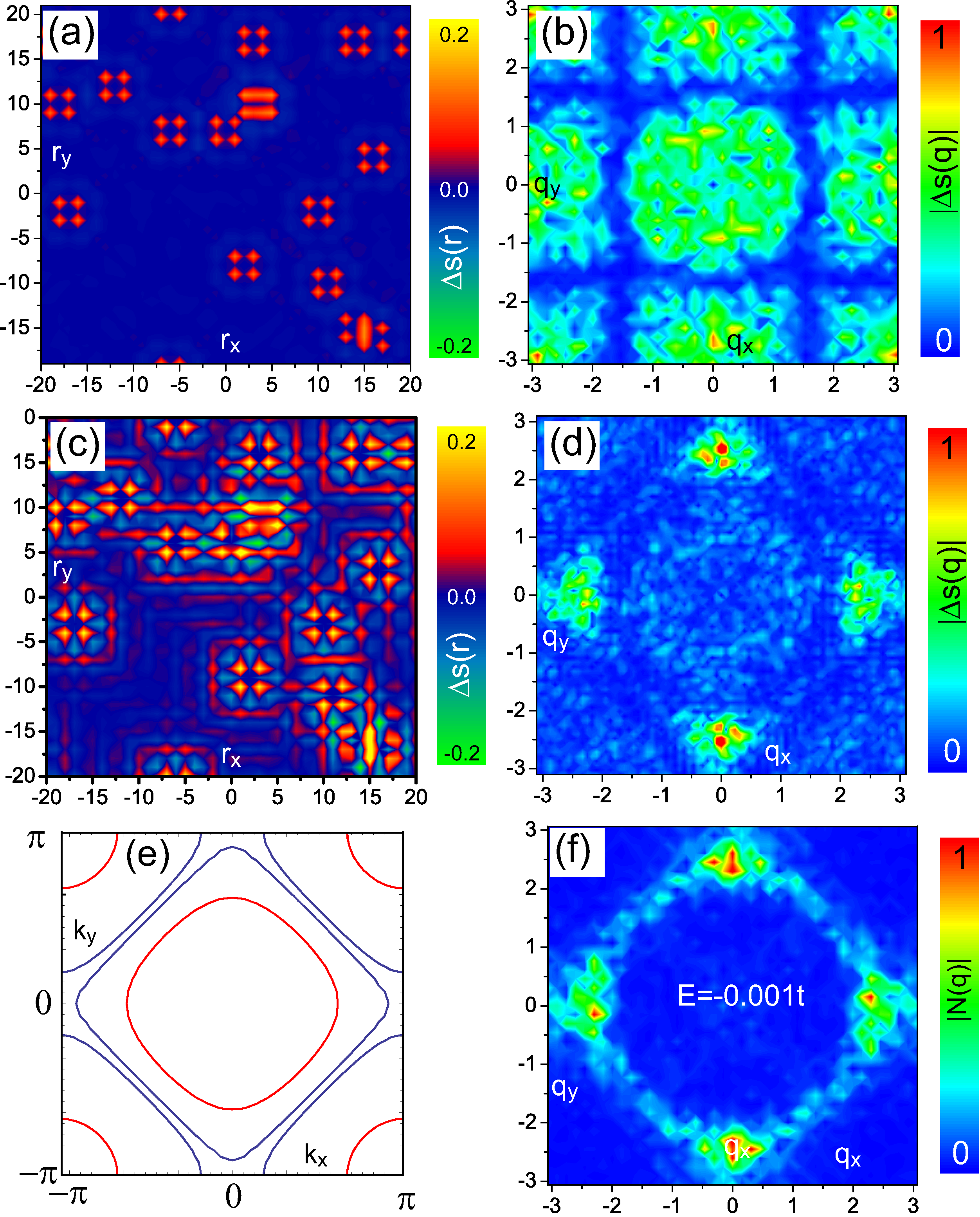}
\caption{(Color online) Kondo lattice with $1\%$ of vacancies. Contour plot of (a) $\Delta s(\bf{r})$, and (b) $|\Delta s(\bf{q})|$ at $T=T_1=0.00002t$. (c) $\Delta s(\bf{r})$, and (d) $|\Delta s(\bf{q})|$ at $T=T_2=0.00022t$. (e) Equal energy contours at $\pm k_B T_{1}$ (blue lines) and $\pm k_B T_{2}$ (red lines). (f) QPI spectrum for $E=-0.001t$ at $T=0.00022t$. }
\label{fig:j147}
\end{figure}
In Figs.~\ref{fig:j147}(a) and \ref{fig:j147}(b) we present the resulting $\Delta s(\bf{r})$ and $|\Delta s(\bf{q})|$, respectively, at $T_1=0.00002t$, which are similar to the results shown in Fig.~\ref{fig:j2u0}. However, upon raising the temperature to $T_2=0.00022t$, we find a qualitative change in the spatial form of $\Delta s({\bf r})$ [see Fig.~\ref{fig:j147}(c)]: The spatial extent of the oscillations as well as their overall amplitude increases significantly. As a result, there also occurs a  significant redistribution of spectral weight in $|\Delta s({\bf q})|$ [see Fig.~\ref{fig:j147}(d)], which is now dominated by four peaks along the bond directions. The enhanced oscillations in the hybridization are now directly reflected in the QPI spectra,  as shown in Fig.~\ref{fig:j147}(f) where we present the QPI pattern for $E=-0.001t$ which exhibits a very similar structure to $|\Delta s({\bf q})|$. We note that these drastic changes in the hybridization and the QPI spectra occur well below the Kondo temperature $T_K$, which we estimated from the vanishing of the hybridization for a clean system as $T_K \approx 0.004t$, such that $T_1<T_2 \ll T_K$. To understand these significant changes, it is necessary to consider the states in the Brillouin zone (BZ), which are excited at $T_{1,2}$, and which (roughly) lie between the equal energy contours $E=\pm k_B T_{1,2}$ shown in Fig.~\ref{fig:j147}(e). At $T_{1}$, states in only a small region of the BZ  are excited [between the blue lines in Fig.~\ref{fig:j147}(e)], leading to the pattern of $\Delta s({\bf r})$ shown in Fig.~\ref{fig:j147}(a). However, at $T_{2}$, states in a much larger portion of the BZ are excited (between the red lines), leading to the significant changes in $\Delta s$ and the QPI spectra shown in Figs.~\ref{fig:j147}(c), \ref{fig:j147}(d), and \ref{fig:j147}(f), respectively. The strong temperature dependence discussed here is mainly an effect of the weak $f$-electron dispersion, resulting from a small value of $I$, and will therefore decrease with increasing strength of the antiferromagnetic interactions.

Our results discussed above possess two important experimental implications. First, it is apparent from the contour plots of $\Delta s$ in Figs.~\ref{fig:j2u-0.7}(a) and \ref{fig:j147}(c) that already a small concentration of defects, indeed as small as 1\%, can essentially disorder the hybridization in the entire system. To quantify this, we consider the hybridization $s({\bf r})$ at a site ${\bf r}$ disordered when it deviates by more than 1\% from its value in the clean system (such a deviation corresponds to the experimental resolution limit in measuring the resulting changes in the energy width of the Kondo resonance \cite{All13}). We then find that for the case of nonmagnetic defects shown in Fig.~\ref{fig:j2u-0.7}(a),  57.9\% of the sites are disordered, whereas for the case shown in Fig.~\ref{fig:j147}(c), 78.2\% of the sites are disordered. This result provides an explanation for the strong disorder effects observed by Hamidian {\it et al.}\cite{Hamidian-11} in (weakly) 1\% Th-doped URu$_2$Si$_2$.

Second, while the hybridization can be strongly disordered even for a small concentration of defects, the specific heat of the system is hardly affected. The specific heat is proportional to the spatially averaged density of states, \cite{Hewson} $\langle N_{tot}(r,\omega) \rangle = \langle N_{c}(r,\omega) \rangle + \langle N_{f}(r,\omega) \rangle$ at $\omega=0$, where $N_{c}, N_{f}$ are the density of states of the $c$- and $f$-quasiparticle bands, \cite{Yuan-12} respectively, and $\langle ...\rangle$ denotes spatial averaging.  For the two most disordered cases shown in Figs.~\ref{fig:j2u-0.7}(a) and \ref{fig:j147}(c), the specific heat decreases only by about 8\% for the system in Fig.~\ref{fig:j2u-0.7}(a), while the change is less than 0.1\% for the case of Fig.~\ref{fig:j147}(c). These changes are quantitatively consistent with the changes seen experimentally in the specific heat of heavy-fermion materials with defect concentration of 1\%.~\cite{Pikul} Thus, we conclude that while the electronic structure of heavy-fermion materials can be heavily disordered already by small concentration of defects, the disorder's main signature appears in the hybridization and $dI/dV$ patterns, while the spatially averaged specific heat undergoes only modest changes, thus explaining the apparent contradiction between spectroscopy and thermodynamics.

{\it Acknowledgments.}
We would like to thank J.~C.~Davis, F.~Steglich, S.~Wirth, and A.~Yazdani for stimulating discussions.
This work is supported by the U.S.
Department of Energy under Award No. DE-FG02-05ER46225 (D.K.M.).

\end{document}